\documentclass[conference]{IEEEtran}
\IEEEoverridecommandlockouts
\usepackage{cite}
\usepackage{amsmath,amssymb,amsfonts}
\usepackage{algorithmic}
\usepackage{graphicx}
\usepackage{textcomp}
\usepackage{xcolor}
\usepackage{hyperref}
\usepackage{placeins}
\usepackage{enumitem}
\usepackage{algorithm}
\usepackage{orcidlink}
\usepackage{subfig}
\hypersetup{
    colorlinks,
    linkcolor={blue!80!black},
    citecolor={blue!80!black},
    urlcolor={blue!80!black}
}
\def\BibTeX{{\rm B\kern-.05em{\sc i\kern-.025em b}\kern-.08em
    T\kern-.1667em\lower.7ex\hbox{E}\kern-.125emX}}
\begin{document}

\title{Playing Dice with the Universe: Programming Quantum Computers to Play Traditional Games\\
%\thanks{TZ was funded by the Natural Sciences and Engineering Research Council of Canada (NSERC) and the Simons Foundation.  VKM was funded by the Simons Foundation.}
}

\author{\IEEEauthorblockN{Tristan Zaborniak \orcidlink{0000-0002-4301-0861}}
\IEEEauthorblockA{\textit{Department of Computer Science} \\
\textit{University of Victoria}\\
Victoria, BC, Canada \\
\href{mailto:tristanz@uvic.ca}{tristanz@uvic.ca}}
\and
\IEEEauthorblockN{Vikram Khipple Mulligan \orcidlink{0000-0001-6038-8922}}
\IEEEauthorblockA{\textit{Center for Computational Biology} \\
\textit{Flatiron Institute}\\
New York, NY, United States of America \\
\href{mailto:vmulligan@flatironinstitute.org}{vmulligan@flatironinstitute.org}}
% \and
% \and
% \and
% \and
}

\maketitle

\begin{abstract}
    The challenge of programming classical computers to play traditional, competitive games against human players has helped to advance classical hardware and software.  Quantum computers have the potential to play games in a unique way: programmed \textit{only} with the rules of a game, they should be able to implicitly represent all future paths of a game leading to wins, losses, or draws, and to sample from this path set to identify moves that maximize the likelihood of a win.  This permits skilled play without hard-coded or machine-learned strategy.  As a proof of principle, we present early results obtained after programming the D-Wave quantum annealer with the rules of tic-tac-toe, enabling it to play against a human or classical computer opponent.  We anticipate that, as it has for classical computers, game-playing will serve as an important real-world benchmark for quantum computers.
\end{abstract}

\begin{IEEEkeywords}
    quantum annealing, games, tic-tac-toe, noughts and crosses, game theory
\end{IEEEkeywords}
    
\section{Introduction}
\label{sec_introduction}

    Quantum computing has incredible potential for solving currently-intractable optimization problems underlying challenges in biology, chemistry, physics, and engineering.
    %In past work, we have demonstrated applications in peptide design \cite{mulligan_designing_2019} and in molecular docking \cite{pandey_multibody_2022}.
    However, advancement of the field suffers since many applications seem esoteric to laypeople.  
    Classical computing once suffered the same problem.  A turning point came in 1950, when University of Toronto graduate student Joseph Kates
    introduced ``Bertie the Brain'' at the Canadian National Exhibition (CNE).  ``Bertie'' was the first computer to play a well-known game, tic-tac-toe, serving as an easily understandable way to introduce laypeople to classical computers' potential to perform tasks involving complicated reasoning.  Although popular with CNE crowds, the paradigm-shifting significance of ``Bertie'' was not recognized at the time, warranting only the briefest of mentions in a \textit{Toronto Star} humour column \cite{sinclair_radio_1950}.  
    
    Three years later, in 1953, Alan Turing published a speculative article about the potential for computers to play chess \cite{turing_digital_1953}.  By 1955, machine learning (ML) allowed computers to play checkers \cite{samuel1959some}, launching serious attempts to develop skilful game-playing computers.  By 1996, IBM's specialized Deep Blue computer was able to beat grandmaster Garry Kasparov in chess \cite{kasparov1996day}, and in 2016, Google's AlphaGo neural net, running on widely-available cloud computing hardware, defeated Go grandmaster Lee Sedol \cite{silver2017mastering} --- a major milestone in the current deep learning renaissance.  
    
    As a side-effect of research into developing game-playing computers, games involving play against computer opponents have become common pastimes, and this fuelled the rise of consumer graphics processing units (GPUs), which in turn proved invaluable for scientific computing and deep learning.  Thus, over the last 76 years, the task of programming classical computers to play games has propelled both methodological advancements in classical computing and public engagement with classical computers, with myriad unanticipated benefits.

    Specialized quantum games, with gameplay that incorporates the concepts of superposition, entanglement, and measurement triggering wavefunction collapse, have been created to teach humans about quantum mechanics and quantum computing \cite{eisert1999quantum,goff2002quantum,goff2006quantum}. Quantum algorithms also exist to solve single-player \textit{puzzles} such as Sudoku \cite{mucke2024simple}.  However, the authors are not aware of major forays into developing quantum algorithms to play \textit{traditional}, \textit{competitive} games.  Taking, for example, chess and Go, whose decision trees are astronomically vast, there is potential for quantum advantage, given that quantum computers has been shown to offer advantages over classical computers for solving hard combinatorial problems in principle \cite{superpolynomial, Hartung2025} and in practice \cite{MunozBauza2025, Shaydulin2024}. 
    
    Classical computers have a well-defined state at every point in a calculation, limiting them to explicit consideration of a relatively small number of possible future paths of a game at a time. As a result, classical game-playing is largely limited to some combination of executing fixed algorithms with hard-coded strategies for winning a particular game, or using ML-based study of past instances of a particular game to determine future gameplay. Quantum computers, in contrast, may exist in superpositions of astronomically many states, permitting implicit representation of the full decision tree from a given game state. A quantum computer programmed \emph{only} with the rules and objective of a game could, in principle, sample next moves preferentially from paths that maximize win probability \emph{without} prior knowledge of optimal strategy or typical gameplay.
        
    To explore this idea, we programmed the D-Wave Advantage quantum annealer to play tic-tac-toe.  A ``quantum'' variant of this game exists, with altered rules in which each player makes simultaneous pairs of entangled moves to create an increasingly entangled game \textit{history} \cite{goff2002quantum,goff2006quantum}.  However, our focus was on having the quantum computer play the \textit{traditional} tic-tac-toe game by modelling all \textit{future} paths of the game as a superposition of states to choose optimal moves.  While this is a task mastered by classical computers 76 years ago, we constrained ourselves to providing the quantum annealer with \emph{only} information about the rules of gameplay and the win conditions, including no strategy information beyond the assumption that a player with a win available on their move would take that win (a general assumption for competitive games of complete and perfect information).  Although this is a very early foray into developing game-playing algorithms for quantum computers, we demonstrate a working tic-tac-toe playing quantum algorithm executing on current hardware, the D-Wave Advantage 2 quantum annealer, and show that it is able to win consistently against a randomly-playing opponent, and often to force a draw against a ``perfect'' opponent.  We discuss preliminary findings, potential avenues for improvement, and generalization to other competitive games of perfect or imperfect information.

\section{Methods}
\label{sec_methods}

    In order to model the entire space of future moves of a game from a given current state, a quantum computer's state space must be at least as large as the game's state space.  Tic-tac-toe is a convenient proof of principle, since (ignoring board symmetry) there are at most $9!=362,880$ possible games.  Thus, 19 qubits (yielding a state space of 524,288 states) are needed at a minimum.  In practice, many games lead to an early win, so only 255,168 games are actually possible \cite{juul_255168_2003}.  A very clever algorithm might shave off one qubit; however, we do not seek to be that miserly, or that clever.
    
    We chose to carry out our quantum computations with the D-Wave quantum annealer for a proof of principle due to the optimization-like nature of the problem of sampling from a desired distribution of game states, and due to the maturity of the D-Wave hardware.
    For initial experiments playing against random opponents, we used D-Wave's default linear transition from the starting to Ising-like ending Hamiltonian \cite{johnson2011quantum}.  For later games against a classical ``perfect'' player, we implemented an inhomogeneous driving scheme to improve likelihood of sampling ground states \cite{adame2020inhomogeneous}.
    Other driving schemes could also be applied; alternatively, our approach could be used with the quantum approximate optimization algorithm (QAOA) \cite{farhi_quantum_2014}, quantum imaginary-time evolution (QITE) \cite{motta_determining_2020}, or other gate-based optimization algorithms.  
    
    A game of tic-tac-toe is most nine moves long. We define up to nine ``move registers'', each representing one move, of nine qubits each.  These represent all possible games of tic-tac-toe using a one-hot encoding indicating the position filled by an \texttt{X} or an \texttt{O} in a given move.  As the game progresses, the setup will be altered so that the superposition of states represents the subset of possible games that are compatible with the current state of play.  Qubits in the ``move registers'' will be designated $m_{i,x,y}$ where $i \in \left[ 1,9 \right]$ represents the move index and $x \in \left[ 1,3 \right]$ and $y \in \left[ 1,3 \right]$ are the coordinates of the grid cell.  We also define a ``line register'' of five qubits, designated $l_i$, with $i \in \left[5,9\right]$.  These are used to detect the presence of three of the same symbol in a row on the $i^\text{th}$ move.  (Note that it is not possible to have three in a row before the fifth move).  A five-qubit ``no-line register'', designated $n_i$, complements the line register by ensuring that line register qubits remain in the 0 state when no line of three is completed on a given turn.  We also designate a three-qubit ``existing-line register''.  These qubits, designated $e_i$ for $i \in \left[7,9\right]$, are used to indicate that at least one previous move has resulted in three in a row.  Finally, we designate an additional ``win register'' of five qubits, designated $w_i$, with $i \in \left[5,9\right]$.  At most, one of these will be set to 1 to indicate that the game was won on a particular move.  Ancillary qubits described below are also needed to connect these registers.

    The quantum annealer is programmed by setting \emph{biases} $h_i$, each giving a bonus or penalty for a qubit with index $i$ being in the 1 state, and \emph{couplings} $J_{i_1,i_2}$, each giving a bonus or penalty for qubits with indices $i_1$ and $i_2$ both being in the 1 state.  For the subsequent steps, we define \texttt{AND}, \texttt{OR}, \texttt{WNOT}, \texttt{WAND}, and \texttt{PW} gates, to be used in encoding the rules and objectives of tic-tac-toe (subsections \ref{subsec_and_gate}, \ref{subsec_or_gate}, \ref{subsec_weak_not_gate}, \ref{subsec_weak_and_gate}, and \ref{subsec_pw_gate}).  We add biases and couplings to encode the allowed moves in tic-tac-toe (subsection \ref{subsec_tictactoe_rules}) and to recognize wins (subsection \ref{subsec_tictactoe_objective}).  Finally, we define a simple, classical outer algorithm using quantum annealing-based sampling statistics to choose the move maximizing win probability (subsection \ref{subsec_tictactoe_nextmove}).

\subsection{Defining an \texttt{AND} gate}
\label{subsec_and_gate}

    An $\texttt{AND}\left(i_1,i_2,o,p_\texttt{AND}\right)$ gate couples three qubits (inputs $i_1$ and $i_2$; output $o$) using a penalty $p_\texttt{AND}$, yielding the overall penalty values given in Table \ref{table_and_truthtable}.  Like a classical \texttt{AND} gate, the degenerate ground state (bold in Table \ref{table_and_truthtable}) includes all states in which at least one input is 0 and the output is 0, and the state in which both inputs are 1 and the output is 1.  This is accomplished by adding $p_\texttt{AND}$ to $J_{i_1,i_2}$, adding $3 p_\texttt{AND}$ to $h_{o}$, and adding $-2 p_\texttt{AND}$ to each of $J_{i_1,o}$ and $J_{i_2,o}$.

    \begin{table}
    \centering
    \caption{Penalty value tables for (a) \texttt{AND}, (b) \texttt{OR}, and (c) \texttt{WNOT} gates.}
    \subfloat[\scriptsize{$\texttt{AND}\left(i_1,i_2,o,p_\texttt{AND}\right)$}]{

            %\caption{Penalty value table for $\texttt{AND}\left(i_1,i_2,o,p_\texttt{AND}\right)$ gate.}
            \begin{tabular}{p{0.5mm} p{0.5mm} p{0.5mm}|p{7mm}}
                \scriptsize{$i_1$} & \scriptsize{$i_2$} & \scriptsize{$o$} & \begin{centering} \scriptsize{Penalty values} \end{centering} \\
                \hline
                \scriptsize{\textbf{0}} & \scriptsize{\textbf{0}} & \scriptsize{\textbf{0}} & \scriptsize{\textbf{0}} \\
                \scriptsize{\textbf{1}} & \scriptsize{\textbf{0}} & \scriptsize{\textbf{0}} & \scriptsize{\textbf{0}} \\
                \scriptsize{\textbf{0}} & \scriptsize{\textbf{1}} & \scriptsize{\textbf{0}} & \scriptsize{\textbf{0}} \\
                \scriptsize{\textbf{1}} & \scriptsize{\textbf{1}} & \scriptsize{\textbf{1}} & \scriptsize{\textbf{0}} \\
                \scriptsize{1} & \scriptsize{1} & \scriptsize{0} & \scriptsize{$p_\texttt{AND}$} \\
                \scriptsize{0} & \scriptsize{0} & \scriptsize{1} & \scriptsize{$3 p_\texttt{AND}$} \\
                \scriptsize{1} & \scriptsize{0} & \scriptsize{1} & \scriptsize{$p_\texttt{AND}$} \\
                \scriptsize{0} & \scriptsize{1} & \scriptsize{1} & \scriptsize{$p_\texttt{AND}$}
            \end{tabular}
            \label{table_and_truthtable}
        %\end{table}
    }
    \quad
    \subfloat[\scriptsize{$\texttt{OR}\left(i_1,i_2,o,p_\texttt{OR}\right)$}]{
        %\begin{table}
            %\centering
            %\caption{Penalty value table for $\texttt{OR}\left(i_1,i_2,o,p_\texttt{OR}\right)$ gate.}
            \begin{tabular}{p{0.5mm} p{0.5mm} p{0.5mm}|p{7mm}}
                \scriptsize{$i_1$} & \scriptsize{$i_2$} & \scriptsize{$o$} & \begin{centering} \scriptsize{Penalty values} \end{centering} \\
                \hline
                \scriptsize{\textbf{0}} & \scriptsize{\textbf{0}} & \scriptsize{\textbf{0}} & \scriptsize{\textbf{0}} \\
                \scriptsize{\textbf{1}} & \scriptsize{\textbf{0}} & \scriptsize{\textbf{1}} & \scriptsize{\textbf{0}} \\
                \scriptsize{\textbf{0}} & \scriptsize{\textbf{1}} & \scriptsize{\textbf{1}} & \scriptsize{\textbf{0}} \\
                \scriptsize{\textbf{1}} & \scriptsize{\textbf{1}} & \scriptsize{\textbf{1}} & \scriptsize{\textbf{0}} \\
                \scriptsize{1} & \scriptsize{0} & \scriptsize{0} & \scriptsize{$p_\texttt{OR}$} \\
                \scriptsize{0} & \scriptsize{1} & \scriptsize{0} & \scriptsize{$p_\texttt{OR}$} \\
                \scriptsize{0} & \scriptsize{0} & \scriptsize{1} & \scriptsize{$p_\texttt{OR}$} \\
                \scriptsize{1} & \scriptsize{1} & \scriptsize{0} & \scriptsize{$3p_\texttt{OR}$} \\
            \end{tabular}
            \label{table_or_truthtable}
        %\end{table}
    }
    \quad
    \subfloat[\scriptsize{$\texttt{WNOT}\left(i,o,p_\texttt{WN}\right)$}]{
        \begin{tabular}{p{0.5mm} p{0.5mm}|p{7mm}}
            \scriptsize{$i$} & \scriptsize{$o$} & \begin{centering} \scriptsize{Penalty values} \end{centering} \\
            \hline
            \scriptsize{\textbf{0}} & \scriptsize{\textbf{0}} & \scriptsize{\textbf{0}} \\
            \scriptsize{\textbf{0}} & \scriptsize{\textbf{1}} & \scriptsize{\textbf{0}} \\
            \scriptsize{\textbf{1}} & \scriptsize{\textbf{0}} & \scriptsize{\textbf{0}} \\
            \scriptsize{1} & \scriptsize{1} & \scriptsize{$p_\texttt{WN}$} \\
        \end{tabular}
        \label{table_wnot_truthtable}
    }
    \end{table}

\subsection{Defining an \texttt{OR} gate}
\label{subsec_or_gate}

    An $\texttt{OR}\left(i_1,i_2,o,p_\texttt{OR}\right)$ gate may be created by setting $h_{i_1}=h_{i_2}=h_{o}=J_{i_1,i_2}=p_\texttt{OR}$, and $J_{i_1,o}=J_{i_2,o}=-2p_\texttt{OR}$.  If any of these biases or couplings has already been set, we add to them.  The gate's penalties, shown in Table \ref{table_or_truthtable}, establish a ground state (bold) analogous to a classical \texttt{OR} gate.

\subsection{Defining a \texttt{WNOT} (``weak not'') gate}
\label{subsec_weak_not_gate}

    We define a third gate called $\texttt{WNOT}\left(i,o,p_\texttt{WN}\right)$.  This is similar to a classical \texttt{NOT} gate, but weaker: if the input qubit $i$ is 1, the ground state of the Hamiltonian ensures that the output qubit $o$ is always 0, but if the input is 0, the output qubit can be 0 or 1.  This is accomplished by adding $p_\texttt{WN}$ to $h_i$, nothing to $h_o$, and $-p_\texttt{WN}$ to $J_{i,o}$.  The resulting penalty values are shown in Table \ref{table_wnot_truthtable}.

\subsection{Defining a \texttt{WAND} (``weak \texttt{AND}'') gate}
\label{subsec_weak_and_gate}

    We also define a $\texttt{WAND}\left(i_1,i_2,a,o,p_\texttt{WA}\right)$ gate (or ``weak \texttt{AND}'' gate), which couples two input qubits, $i_1$ and $i_2$, an ancillary qubit $a$, and an output qubit $o$ using a penalty $p_\texttt{WA}$.  In the degenerate ground state (bold in Table \ref{table_weak_and_truthtable}), $o$ is 1 if both $i_1$ and $i_2$ are 1, but $o$ can be either 0 or 1 if either or both of $i_1$ and $i_2$ are 0.  We call this ``\texttt{WAND}'' since its penalty value table resembles, but is distinct from, the truth table for a classical \texttt{AND} gate. In the \texttt{WAND} case, the output is always 1 if the two inputs are 1, but may or may not be 0 otherwise, depending on other couplings.  We accomplish this by starting with an $\texttt{AND}\left(i_1,i_2,a,p_\texttt{WA}\right)$ gate (subsection \ref{subsec_and_gate}), ensuring that in the ground state, $a$ is 1 if and only if $i_1$ and $i_2$ are both 1.  We then couple $a$ and $o$ to ensure that $\left(a,o\right)=\left(1,0\right)$ is penalized, and all other states have energy 0.  We do this by adding $p_\texttt{WA}$ to the bias $h_a$ for $a$, and $-p_\texttt{WA}$ to the coupling $J_{a,o}$ between $a$ and $o$.

    \begin{table}
        \centering
        \caption{Penalty value tables for (a) \texttt{WAND} and (b) \texttt{PW} gates.}
        \subfloat[$\texttt{WAND}\left(i_1,i_2,a,o,p_\texttt{WA}\right)$]{
            \begin{tabular}{c c c c|p{7mm}}
                \scriptsize{$i_1$} & \scriptsize{$i_2$} & \scriptsize{$a$} & \scriptsize{$o$} & \begin{centering} \scriptsize{Penalty values} \end{centering} \\
                \hline
                \scriptsize{\textbf{0}} & \scriptsize{\textbf{0}} & \scriptsize{\textbf{0}} & \scriptsize{\textbf{0}} & \scriptsize{\textbf{0}} \\
                \scriptsize{\textbf{1}} & \scriptsize{\textbf{0}} & \scriptsize{\textbf{0}} & \scriptsize{\textbf{0}} & \scriptsize{\textbf{0}} \\
                \scriptsize{\textbf{0}} & \scriptsize{\textbf{1}} & \scriptsize{\textbf{0}} & \scriptsize{\textbf{0}} & \scriptsize{\textbf{0}} \\
                \scriptsize{\textbf{0}} & \scriptsize{\textbf{0}} & \scriptsize{\textbf{0}} & \scriptsize{\textbf{1}} & \scriptsize{\textbf{0}} \\
                \scriptsize{\textbf{1}} & \scriptsize{\textbf{0}} & \scriptsize{\textbf{0}} & \scriptsize{\textbf{1}} & \scriptsize{\textbf{0}} \\
                \scriptsize{\textbf{0}} & \scriptsize{\textbf{1}} & \scriptsize{\textbf{0}} & \scriptsize{\textbf{1}} & \scriptsize{\textbf{0}} \\
                \scriptsize{\textbf{1}} & \scriptsize{\textbf{1}} & \scriptsize{\textbf{1}} & \scriptsize{\textbf{1}} & \scriptsize{\textbf{0}} \\
                \scriptsize{1} & \scriptsize{1} & \scriptsize{0} & \scriptsize{0} & \scriptsize{$p_\texttt{WA}$} \\
                \scriptsize{0} & \scriptsize{0} & \scriptsize{1} & \scriptsize{0} & \scriptsize{$4 p_\texttt{WA}$} \\
                \scriptsize{1} & \scriptsize{0} & \scriptsize{1} & \scriptsize{0} & \scriptsize{$2 p_\texttt{WA}$} \\
                \scriptsize{0} & \scriptsize{1} & \scriptsize{1} & \scriptsize{0} & \scriptsize{$2 p_\texttt{WA}$} \\
                \scriptsize{1} & \scriptsize{1} & \scriptsize{1} & \scriptsize{0} & \scriptsize{$p_\texttt{WA}$} \\
                \scriptsize{1} & \scriptsize{1} & \scriptsize{0} & \scriptsize{1} & \scriptsize{$p_\texttt{WA}$} \\
                \scriptsize{0} & \scriptsize{0} & \scriptsize{1} & \scriptsize{1} & \scriptsize{$3 p_\texttt{WA}$} \\
                \scriptsize{1} & \scriptsize{0} & \scriptsize{1} & \scriptsize{1} & \scriptsize{$p_\texttt{WA}$} \\
                \scriptsize{0} & \scriptsize{1} & \scriptsize{1} & \scriptsize{1} & \scriptsize{$p_\texttt{WA}$}
            \end{tabular}
            \label{table_weak_and_truthtable}
        }
        \quad
        \subfloat[$\texttt{PW}\left(i_1,i_2,o_1,o_2,p_\texttt{PW}\right)$]{
            \begin{tabular}{c c c c|p{7mm}}
                \scriptsize{$i_1$} & \scriptsize{$i_2$} & \scriptsize{$o_1$} & \scriptsize{$o_2$} & \begin{centering}  \scriptsize{Penalty values} \end{centering} \\
                \hline
                \scriptsize{\textbf{0}} & \scriptsize{\textbf{0}} & \scriptsize{\textbf{0}} & \scriptsize{\textbf{0}} & \scriptsize{\textbf{0}} \\
                \scriptsize{\textbf{0}} & \scriptsize{\textbf{1}} & \scriptsize{\textbf{0}} & \scriptsize{\textbf{1}} & \scriptsize{\textbf{0}} \\
                \scriptsize{\textbf{1}} & \scriptsize{\textbf{0}} & \scriptsize{\textbf{1}} & \scriptsize{\textbf{0}} & \scriptsize{\textbf{0}} \\
                \scriptsize{\textbf{1}} & \scriptsize{\textbf{1}} & \scriptsize{\textbf{1}} & \scriptsize{\textbf{0}} & \scriptsize{\textbf{0}} \\
                \scriptsize{0} & \scriptsize{0} & \scriptsize{0} & \scriptsize{1} & \scriptsize{$p_\texttt{PW}$} \\
                \scriptsize{0} & \scriptsize{0} & \scriptsize{1} & \scriptsize{0} & \scriptsize{$p_\texttt{PW}$} \\
                \scriptsize{0} & \scriptsize{1} & \scriptsize{0} & \scriptsize{0} & \scriptsize{$p_\texttt{PW}$} \\
                \scriptsize{1} & \scriptsize{0} & \scriptsize{0} & \scriptsize{0} & \scriptsize{$p_\texttt{PW}$} \\
                \scriptsize{0} & \scriptsize{0} & \scriptsize{1} & \scriptsize{1} & \scriptsize{$3 p_\texttt{PW}$} \\
                \scriptsize{0} & \scriptsize{1} & \scriptsize{1} & \scriptsize{1} & \scriptsize{$p_\texttt{PW}$} \\
                \scriptsize{1} & \scriptsize{0} & \scriptsize{1} & \scriptsize{1} & \scriptsize{$3 p_\texttt{PW}$} \\
                \scriptsize{1} & \scriptsize{1} & \scriptsize{1} & \scriptsize{1} & \scriptsize{$p_\texttt{PW}$} \\
                \scriptsize{1} & \scriptsize{1} & \scriptsize{0} & \scriptsize{1} & \scriptsize{$2 p_\texttt{PW}$} \\
                \scriptsize{1} & \scriptsize{1} & \scriptsize{0} & \scriptsize{0} & \scriptsize{$2 p_\texttt{PW}$} \\
                \scriptsize{0} & \scriptsize{1} & \scriptsize{1} & \scriptsize{0} & \scriptsize{$p_\texttt{PW}$} \\
                \scriptsize{1} & \scriptsize{0} & \scriptsize{0} & \scriptsize{1} & \scriptsize{$3 p_\texttt{PW}$} \\
            \end{tabular}
            \label{table_pw_truthtable}
        }
    \end{table}

\subsection{Defining a \texttt{PW} (``past win'') gate}
\label{subsec_pw_gate}

    We define a fifth gate called $\texttt{PW}\left(i_1,i_2,o_1,o_2,p_\texttt{PW}\right)$, for ensuring that past wins prevent future instances of three characters in a row from registering as a win.  Its penalty values are shown in Table \ref{table_pw_truthtable}, and are configured by setting $h_{i_1}=h_{i_2}=h_{o_1}=h_{o_2}=J_{o_1,o_2}=J_{i_1,o_2}=p_\texttt{PW}$, $J_{i_1,o_1}=J_{i_2,o_2}=-2p_\texttt{PW}$, and $J_{i_1,i_2}=J_{i_2,o_1}=-p_\texttt{PW}$.  If any of these biases or couplings has already been set, we add to them.

\subsection{Programming allowed moves into the Hamiltonian}
\label{subsec_tictactoe_rules}

    Our first goal was to set biases and couplings to construct a highly degenerate ground state in which bitstrings corresponding to possible allowed games of tic-tac-toe are isoenergetic and lowest-energy, and all other bitstrings are excited states.  Specifically, we encoded the following rules:

    \begin{enumerate}
        \item \textit{Exactly one square must be marked per move.}  
        This necessitates a positive multi-selection penalty $p_\text{ms}$ to be added to the couplings for all pairs of qubits $\left(m_{i,x,y},m_{i,x',y'}\right)~\forall~\left(x,y\right)\ne\left(x',y'\right)$ in each move register $i$, and a bonus of $-\frac{p_\text{ms}}{2}$ to be added to all biases for each qubit $m_{i,x,y}\forall\left(x,y\right)$ in each move register $i$.
        \item \textit{The same square may not be marked twice.}  
        This requires a positive overlap penalty $p_\text{o}$ to be added to the couplings for every pair of qubits $\left(m_{i,x,y}, m_{j,x,y}\right)~\forall~i\neq j$, and a bonus $-\frac{p_\text{o}}{2}$ to be added to each bias for each qubit $m_{i,x,y}$.
        \item \textit{Moves that have been made remain in place and reduce the state space.}  
        When programming the annealer for a mid-game move $n>i$, if an \texttt{X} or \texttt{O} has been placed in a square $\left(x,y\right)$ on move $i$, then a positive penalty of $p_\text{o}$ is added to the biases of all qubits $m_{j,x,y}~\forall~i \neq j$ to ensure that the ground state does not include any move that marks that square again.  This shrinks the ground state to the superposition of possible games remaining.
    \end{enumerate}

    \noindent Once these values are added to the biases and couplings for the move registers, repeated sampling of the ground state will produce a series of bitstrings each representing a single possible future path of the game given only the valid moves.  The bitstrings represent the square marked in each move, expressed as a one-hot encoding in each move register.  Note that we do not need to represent whether a square is occupied by an \texttt{X} or an \texttt{O}, since this may be inferred from whether the move index is odd or even.

\subsection{Programming the objective into the Hamiltonian}
\label{subsec_tictactoe_objective}

    Unbiased sampling from the ground state of the Hamiltonian established thus far should result in all possible games being sampled with equal probability --- a task already difficult to achieve classically without introducing bias.  It is advantageous to be able to sample selectively from those game paths leading to a win, loss, or draw, however, to allow paths to wins to be sampled from even when rare.
    To this end, we next sought to encode a concept of winning or losing, to implicitly permit efficient sampling from the subset of game paths that result in a win for \texttt{X}, or the subset that result in a win for \texttt{O}.  We also sought to ensure that we limited the game paths to the subset in which players presented with a winning move always embraced that move (minimally strategic play).  For this purpose, we aimed to program the line register qubits so that the ground state would encode a 1 if a given move resulted in a win for that player, and a 0 otherwise.  To accomplish this, we encoded the following rules:

    \begin{enumerate}[resume]
        \item \textit{Three \texttt{X} markings or three \texttt{O} markings in a row, in a column, or along a diagonal amount to a win.}  
        We note that a win is not possible before the fifth move of a tic-tac-toe game.  Thus, for each qubit representing square $\left(x,y\right)$ in move register $i \geq 5$, we add a series of $\texttt{WAND}$ gates with coupling strength $p_\text{line}$.  Each couples input qubits $m_{j,x_1,y_1}$ and $m_{i-2,x_2,y_2}$ to output qubit $m_{i,x,y}$, with one such gate for each pair $\left(\left(x_1,y_1\right),\left(x_2,y_2\right)\right)$ that could complete a row, column, or diagonal with $\left(x,y\right)$, and for each $j < \left(i-2\right) | j \text{ mod } 2 = i \text{ mod } 2$.  Additionally, the ancilla qubit from each of these $\texttt{WAND}$ gates and the output qubit $m_{i,x,y}$ are the input to another $\texttt{WAND}$, also with strength $p_\text{line}$, that connects to output line qubit $l_i$.  The effect is to ensure that on the move that a row, column, or diagonal is filled, the line qubit for that move is set to 1. To ensure that the line qubit is zero in the absence of a line of three completed on move $i$, we anti-constrain line qubit $l_i$ and no-line qubit $n_i$ with a positive coupling $p_\text{noline}$.  We ensure that no-line qubit $n_i$ is 1 if there is no line completed on move $i$ by applying a \texttt{WNOT} gate coupling each ancilla qubit from the second set of \texttt{WAND} gates for the win qubit to the no-win qubit.
        %\textcolor{red}{**In addition to line qubits, need antiline qubits.  Need a qubit for move 5 onward that is 1 if all possible lines are NOT formed, 1 or 0 otherwise.  Need it to be anticoupled to the line qubit for that move.  Note that the ancilla qubit for the second WAND, above, is 1 if there's a line.  Need a WNOT between all of these ancilla qubits and an antiline qubit, so that the antiline qubit is 1 if all of these ancilla qubits are 0, and 1 or 0 if they'e not.  WNOT: i-0 o-0, i-0 o-1, i-1 o-0 are the ground state.  Achieved with h\_i = +p\_wnot, J\_io = -p\_wnot, h\_0 = 0**}
        \item \textit{A player who is able to win on a given move will do so (minimally strategic play).} 
        Effectively, this is encoded in the implementation of the previous rule, by \emph{not} adding win checks for all possible ways of winning in a row, column, or diagonal.  Because two of the moves considered are moves $i$ and $i-2$, the win condition requires that the second \texttt{X} or \texttt{O} that can be involved in the win must be placed two moves (one turn for the winning player) before the third \texttt{X} or \texttt{O} involved in the win.  This means that a game path in which an available win is not taken is an excited state, excluded from the ground state ensemble.  Note that this could be extended to any assumptions that one wished to encode about one's opponent's play; however, in the case of tic-tac-toe, encoding too many assumptions about the opponent's play could yield an emergent strategy that approaches the Nash equilibrium, which inevitably results in a draw \cite{kalyan2024tic}.
        \item \textit{Only the first three-in-a-row is a win.}
        To this point, we ensure that three of the same symbol in a row triggers a 1 in the line qubit for the move on which the third was placed; however, there is no notion that the game stops at this point. We must now couple the line qubits with the win qubits to ensure that a win is only registered on the first move that produces three in a row. To this end, we first apply $\texttt{OR}\left(l_5,l_6,e_7,p_\text{ex}\right)$, $\texttt{OR}\left(l_7,e_7,e_8,p_\text{ex}\right)$, and $\texttt{OR}\left(l_8,e_8,e_9,p_\text{ex}\right)$ so that the existing-line qubits are activated by past instances of three in a row. 
        Next, we strongly couple $l_5$ to $w_5$, since three in a row at move 5 is always a win. 
        Finally, we apply $\texttt{PW}\left(l_5,l_6,w_5,w_6,p_\text{win}\right)$, $\texttt{PW}\left(e_7,l_7,w_6,w_7,p_\text{win}\right)$, $\texttt{PW}\left(e_8,l_8,w_7,w_8,p_\text{win}\right)$, and $\texttt{PW}\left(e_9,l_9,w_8,w_9,p_\text{win}\right)$ gates to ensure that only the first three-in-a-row activates the corresponding win qubit.
        \item \textit{We wish to sample from games in which \texttt{X} wins, or games in which \texttt{O} wins, or games ending in draws.} In order to achieve this, the degenerate ground state must be split based on which player wins.  To sample from states in which \texttt{X} wins, we add a positive penalty $p_\text{wb}$ (for ``win bias'') to the biases of the win qubits for moves on which \texttt{O} could win, and a bonus $-p_\text{wb}$ to the biases for the win qubits representing \texttt{X} winning.  Inverting these penalties permits the opposite: sampling from paths that give \texttt{O} the win.  Adding the positive $p_\text{wb}$ penalty to win qubits for \texttt{X} \textit{and} \texttt{O} results in sampling from the states representing games that are draws.
    \end{enumerate}

    \noindent The above permits tuning of the relative strengths of the various penalties.  For simplicity, we set $p_\text{ms} = p_\text{o} = p_\text{line} = p_\text{noline} = p_\text{ex} = p_\text{win} = p_\text{wb} = 1$.

\subsection{Choosing the next move based on quantum sampling}
\label{subsec_tictactoe_nextmove}

    Given the above setup (or equivalent for another game) encoding the rules and win state, as well as any assumptions about an opponent's strategy (here, the assumption of minimally strategic play), the general algorithm is:

    \begin{enumerate}
        \item Using a quantum optimization algorithm, draw ground-state samples for moves that lead to a win, loss, or draw, carrying out a constant number of samples for each.
        \item Given a set of sampled possible next moves $M$, for each $m \in M$, classically count $n_{\text{win},m}$, $n_{\text{loss},m}$, and $n_{\text{draw},m}$ (the number of future game paths that were sampled that lead to a win, loss or draw for each candidate move $m$, where wins, losses, and draws are inferred from the state of the win qubits), as well as $n_{\text{tot},m} = n_{\text{win},m} + n_{\text{loss},m} + n_{\text{draw},m}$ and $n_\text{tot} = \sum\limits_{m \in M } n_{\text{tot},m}$.
        \item For each $m \in M$, classically compute $\mathcal{P}^{\prime}(\text{win}|m)$ as described below in Eq. \eqref{eq_probability_prime_of_win}.
        \item Classically choose $m$ that maximizes $\mathcal{P}^{\prime}(\text{win}|m)$ as the next move.
    \end{enumerate}

Our goal is to choose the next move $m$ that maximizes $\mathcal{P}(\text{win}|m)$, the probability of winning given the move, which we must classically estimate for each possible next move from the finite set of samples given by the quantum computer.  We first apply Bayes' theorem:

    \begin{equation}
    \mathcal{P}\left(\text{win}|m\right) = \frac {\mathcal{P}\left(m|\text{win}\right) \mathcal{P}\left(\text{win}\right) }{ \mathcal{P}\left(m\right) }
    \label{eq_probability_of_win_1_bayes}
    \end{equation}

\noindent Defining $n_\text{win}=\sum\limits_{m \in M} n_{\text{win},m}$, we substitute into the above $\mathcal{P}(m|\text{win})=n_{\text{win},m}/n_{\text{win}}$ and $\mathcal{P}(m)=n_{\text{tot},m}/n_\text{tot}$ to get:

    \begin{equation}
    \mathcal{P}\left(\text{win}|m\right) = \frac { n_{\text{win},m} n_\text{tot} \mathcal{P}\left(\text{win}\right) }{ n_{\text{win}} n_{\text{tot},m}}
    \label{eq_probability_of_win_2_substitutions}
    \end{equation}

\noindent We observe that $n_\text{tot}$, $\mathcal{P}(\text{win})$, and $n_\text{win}$ are positive and do not vary with $m$, meaning that:

    \begin{equation}
    \mathcal{P}\left(\text{win}|m\right) \propto \frac { n_{\text{win},m} }{ n_{\text{tot},m}}
    \label{eq_probability_of_win_3_proportional}
    \end{equation}

\noindent This permits us to define $\mathcal{P}^{\prime}(\text{win}|m)$, the quantity that, when maximized by the choice of move, also maximizes $\mathcal{P}(\text{win}|m)$:

    \begin{equation}
    \mathcal{P}^{\prime}\left(\text{win}|m\right)=
    \begin{cases}
        0 & \text{if}~n_{\text{tot},m}= 0 \\
        \frac{n_{\text{win},m}}{n_{\text{tot},m}} & \text{otherwise}
    \end{cases}
    \label{eq_probability_prime_of_win}
    \end{equation}

Note that in Eq. \eqref{eq_probability_prime_of_win}, the values of $n_{\text{win},m}$, $n_{\text{loss},m}$, and $n_{\text{draw},m}$ that are summed to determine $n_{\text{tot},m}$ come from sampling three different Hamiltonians.  Each is constructed to yield samples that are unbiased with respect to the choice of next move, so that even if any yields an over- or under-estimate of the overall probability of a win, loss, or draw across \textit{all} moves, the \textit{rank order} of \textit{particular} moves leading to wins is unchanged (as we see going from Eq. \eqref{eq_probability_of_win_2_substitutions} to Eq. \eqref{eq_probability_of_win_3_proportional}).

\subsection{Quantum annealing hyperparameters}

We carried out annealing runs on the DWave Advantage 2 System 1.13 (human or random opponent) or System 1 (perfect opponent), performing 5 sets of 5,000 samples to sample from the set of future game paths leading to each of wins, losses, and draws (for a total of 75,000 samples per move; see section \ref{subsec_tictactoe_objective}, rule 7, and section \ref{subsec_tictactoe_nextmove}).  Minor embeddings for each set of 5,000 samples were found with \texttt{minorminer} \cite{cai_practical_2014}, using 2,048 attempts (random) or 8,192 attempts (perfect), chainlength patience of 128 (random) or 512 (perfect), timeouts of 1,800 s (random) or 3,600 s (perfect), and a tolerance of 256 (random) or 512 (perfect) failed iterations per attempt.

\subsection{Code availability}
\label{subsec_code_availability}

    We implemented our tic-tac-toe-playing quantum computer proof-of-principle using the D-Wave Ocean Python interface.  Our Python code is released under a permissive free-and-open-source MIT licence, and is available from \href{https://github.com/flatironinstitute/quantum_tic_tac_toe_public}{https://github.com/flatironinstitute/quantum\_tic\_tac\_toe\_public}.
    
%\FloatBarrier
\section{Early Results and Discussion}
    
    Against the algorithm described in Section \ref{sec_methods} executed on the D-Wave Advantage 2, we pitted human opponents, as well as classical computer opponents playing randomly or according to Newell's ``perfect'' play algorithm \cite{newell_human_2019}.  The quantum player played each classical computer opponent 60 times, with each opponent starting 30 times.  Across these games, the average number of logical qubits needed for modelling the future game path from the $i^\text{th}$ move, including ancillary qubits, is shown in Table \ref{table_qubits_needed}.  In later moves, there is some variation that arises depending on the past moves chosen.  The average number of physical qubits to which this was mapped on the D-Wave Advantage 2 system using the \texttt{minorminer} algorithm is also shown.

\begin{table}
    \centering
    \caption{Logical and physical qubits used for our implementation.}
    \begin{tabular}{c | c c }
        \scriptsize{Move} & \scriptsize{Avg. logical qubits (range)} & \scriptsize{Avg. physical qubits (range)} \\
        \hline
        \scriptsize{1} & \scriptsize{963.0 (963-963)} & \scriptsize{3167.2 (2907-3608)} \\
        \scriptsize{2} & \scriptsize{492.2 (402-534)} & \scriptsize{1528.5(1154-1895)} \\
        \scriptsize{3} & \scriptsize{241.3 (206-293)} & \scriptsize{721.8 (560-1086)} \\
        \scriptsize{4} & \scriptsize{127.5 (100-156)} & \scriptsize{347.5 (257-450)} \\
        \scriptsize{5} & \scriptsize{86.2 (76-97)} & \scriptsize{214.1 (177-278)} \\
        \scriptsize{6} & \scriptsize{59.6 (55-63)} & \scriptsize{124.7 (107-149)} \\
        \scriptsize{7} & \scriptsize{45.4 (43-47)} & \scriptsize{76.7 (69-87)} \\
        \scriptsize{8} & \scriptsize{33.0 (33-33)} & \scriptsize{48.6 (47-51)} \\
        \scriptsize{9 (trivial case)} & \scriptsize{23.0 (23-23)} & \scriptsize{30.5 (30-32)}
    \end{tabular}
    \label{table_qubits_needed}
\end{table}

\begin{figure}
    \centering
    \includegraphics[]{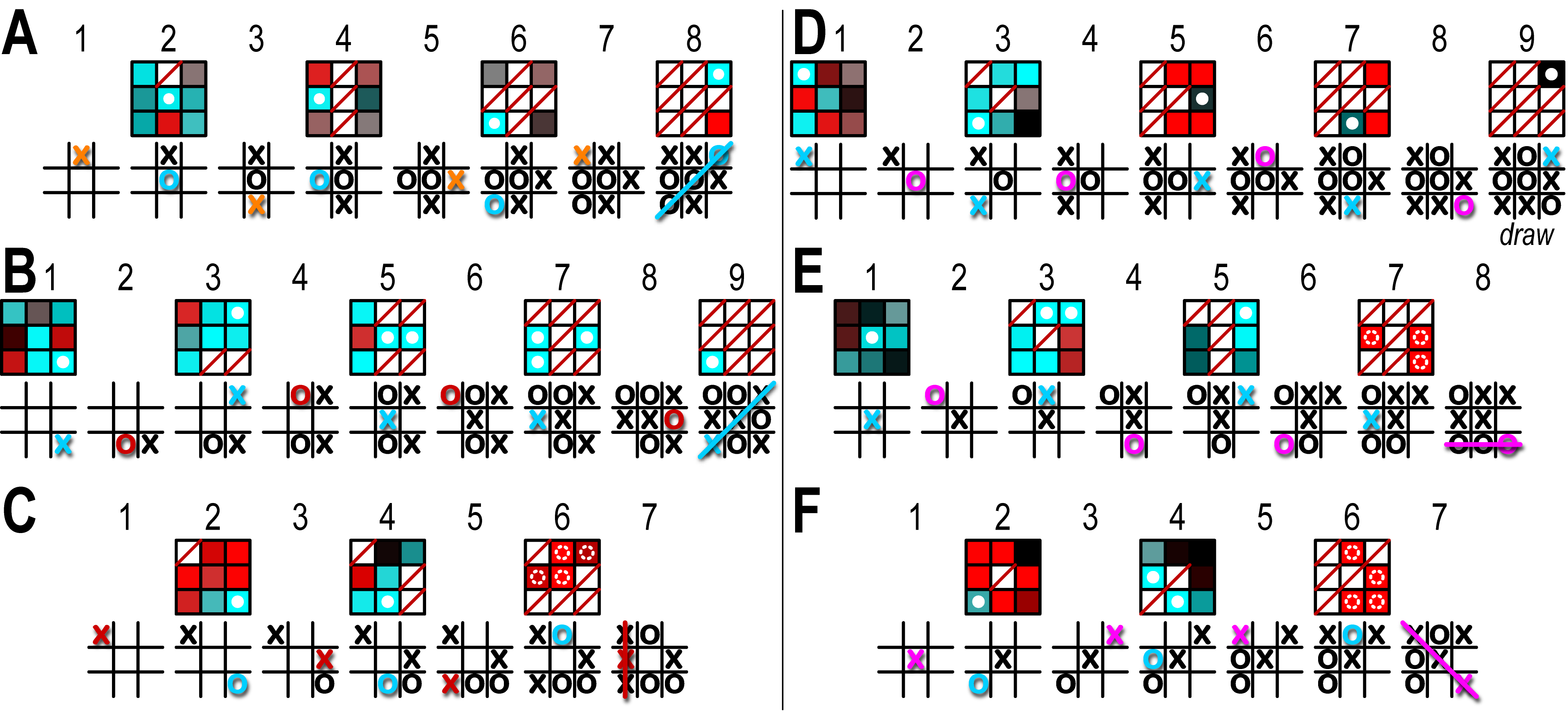}
    \caption{Representative games played by the quantum annealer (cyan \texttt{X}/\texttt{O} marks).  Top grids show estimated $P^{\prime}\left(\text{win}|m\right)$ (cyan intensity) and $P^{\prime}\left(\text{loss}|m\right)$ (red intensity) for candidate moves from quantum sampling.
    %These are expected to correlate with win and loss probabilities by move, respectively.
    Darker squares have higher draw probability. White dots indicate move(s) with the highest win probability.  Dashed white circles indicate that wins are impossible on any move.   (\textbf{A}) The quantum player playing as \texttt{O} appeared to play strategically against a human (orange), fully exploiting the human's mistake in move 3 to give the human a dilemma between two ways of losing after move 6. (\textbf{B})  In a game against a randomly-playing computer opponent as \texttt{O} (red), the quantum player chose moves that would inevitably lead to a win, but which happened to prolong the game. (\textbf{C}) In a rare game that the quantum player playing as \texttt{O} lost to a random player, the lack of any moves with nonzero win probability (dashed white circles) prevented the quantum player from choosing to block the opponent's imminent win to yield a draw.  We subsequently added a preference for draws to losses. (\textbf{D}) Playing as \texttt{X} against a ``perfect'' computer opponent (magenta), the quantum player forced a draw (the best possible outcome) in 28 of 30 games. (\textbf{E}) In 2 of 30 games against the quantum player as \texttt{X}, the ``perfect'' player was able to win by forcing a dilemma. (\textbf{F}) In all games between a ``perfect'' player starting as \texttt{X}, the ``perfect'' player was able to win by exploiting the quantum player's ``optimism'' that wins were possible, and its preference to try to win over forcing a draw.}
    \label{fig:example_games}
\end{figure}

    Qualitatively, the quantum opponent showed somewhat uneven play, sometimes feeling like a real player capable of setting traps and thinking strategically, and at other times missing obvious moves.  \textbf{Fig. \ref{fig:example_games}A} shows a representative game against a human in which the quantum player exploited the human's mistake in move 3 to set the game on a path leading to an inevitable win for the quantum opponent.  The inset for move 4 at top shows that the probability estimates provided by the quantum annealer highlighted the edge positions as far better moves than the corner positions.  However, given the horizontally and vertically symmetric state of the board after move 3, the probability estimates should also have been horizontally and vertically symmetric: the combination of the asymmetric embedding and the stochastic sampling introduced asymmetries.  The same was true in move 6, when the input state was vertically symmetric but the output probability estimates were not.  In move 8, presented with only two choices, one leading to a definite win and the other to a definite loss, the quantum algorithm was able to accurately estimate win and loss probabilities.  In general, accuracy improved as choices narrowed and the state space (and influence of sampling noise) shrank.

    We found that the quantum opponent was always able to defeat the random opponent when the former made the first move, and was able to win in 26 out of 30 games (87\% of the time) when the random opponent started (Table \ref{table_wins_and_losses}).  No games resulted in a draw.  Nevertheless, some quirks were apparent.  The quantum player had no concept of a quick win being more desirable than a slow one: in the game shown in \textbf{Fig. \ref{fig:example_games}B}, the quantum player had guaranteed a win for itself after move 5, but rather than take the immediate win available to it on move 7, it opted for the longer but equally inevitable path to a win, something that most humans would \textit{not} do.  Because our quantum strategy was configured to maximize win probability and not to prefer draws to losses when playing the random opponent, the quantum player ceased to pursue any particular strategy once a win was impossible, leading to the 4 losses observed, an example of which is shown in \textbf{Fig. \ref{fig:example_games}C}.  In this example, at move 6, no path could possibly lead to \texttt{O} (the quantum player) winning, so the quantum player chose the lowest-index square, triggering a loss instead of a draw.
    
    Prior to trying the quantum player against the ``perfect'' classical player, we adjusted the classical outer part of the algorithm, directing it to choose the move that minimizes $P^{\prime}\left(\text{loss}|m\right)$ when quantum sampling yields $P^{\prime}\left(\text{win}|m\right)=0~\forall~m$.  With this adjustment, the quantum player was able to force a draw in all but 2 out of 30 games (93\%) against the ``perfect'' player when playing as \texttt{X} (\textbf{Fig. \ref{fig:example_games}D}).  In the remaining 2 games (7\%), poorly-estimated win probabilities gave the ``perfect'' player an opportunity to engineer a dilemma and seize the win (\textbf{Fig. \ref{fig:example_games}E})  Since the quantum player always seeks a win if one is possible, the ``perfect'' player still had an advantage when the quantum player was \texttt{O}, and was able to win consistently by exploiting the quantum player's ``hopeful'' play.  In 21 games, the quantum player managed to place two adjacent \texttt{O}s that could have led to a win, but the ``perfect'' player blocked this and created a dillemma forcing a loss. 

\begin{table}
    \centering
    \caption{Performance of a quantum tic-tac-toe player against randomly-playing and ``perfect'' classical opponents.}
    \begin{tabular}{c | c c c }
        \scriptsize{Competitors} & \scriptsize{Quantum win} & \scriptsize{Quantum loss} & \scriptsize{Draw} \\
        \hline
        \scriptsize{Quantum (\texttt{X}) vs. Random (\texttt{O})} & \scriptsize{30(100\%)} & \scriptsize{0(0\%)} & \scriptsize{0(0\%)} \\
        \scriptsize{Random (\texttt{X}) vs. Quantum (\texttt{O})} & \scriptsize{26(87\%)} & \scriptsize{4(13\%)} & \scriptsize{0(0\%)} \\
        \scriptsize{Quantum (\texttt{X}) vs. Perfect (\texttt{O})} & \scriptsize{0(0\%)} & \scriptsize{2(7\%)} & \scriptsize{28(93\%)} \\
        \scriptsize{Perfect (\texttt{X}) vs. Quantum (\texttt{O})} & \scriptsize{0(0\%)} & \scriptsize{30(100\%)} & \scriptsize{0(0\%)}
    \end{tabular}
    \label{table_wins_and_losses}
\end{table}

\begin{figure}
    \centering
    \includegraphics[]{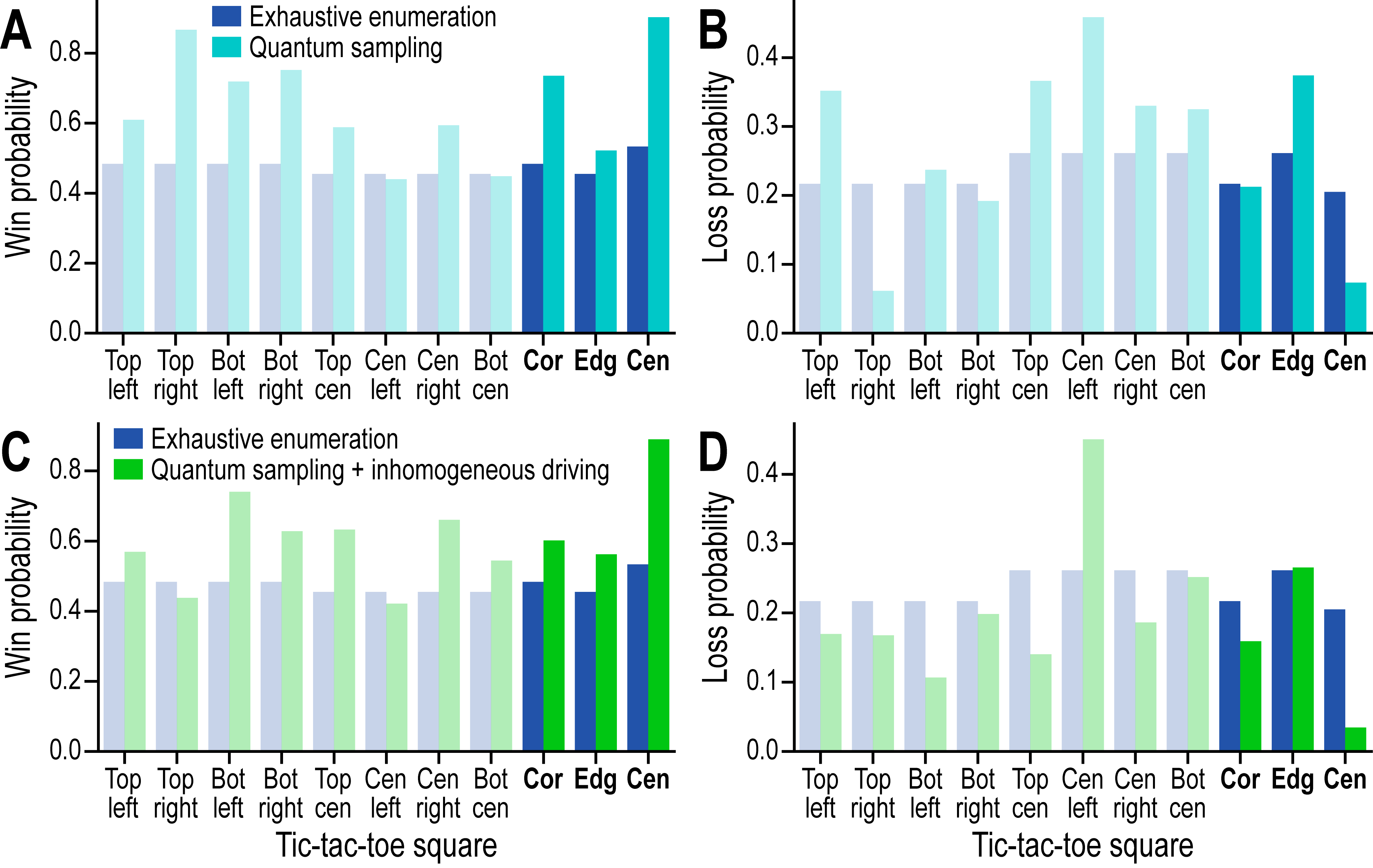}
    \caption{Probability of moves leading to a win (\textbf{A}, \textbf{C}) or loss (\textbf{B}, \textbf{D}), based on exhaustive enumeration of possible games (blue) or quantum sampling without (cyan) or with (green) inhomogeneous driving.  Pale bars show individual corner or edge squares; solid bars represent collective probablities for a corner, edge, or centre square. $P^{\prime}\left(\text{win}|m\right)$ from quantum sampling is expected only to have rank order matching true probability, which was observed.}
    \label{fig:firstmove_analysis}
\end{figure}

    The later games against the ``perfect'' opponent were executed with inhomogeneous driving, while the earlier games were not.  As shown in \textbf{Fig. \ref{fig:firstmove_analysis}A-B}, when the quantum annealer made the first move, win probability estimates were noisy, but it did correctly estimate the rank order of choosing a corner, edge, or centre square.  Inhomogenous driving decreased the noise (\textbf{Fig. \ref{fig:firstmove_analysis}C-D}).  Since $P^{\prime}\left(\text{win}|m\right) \propto P\left(\text{win}|m\right)$, we only expect rank order to match.  However, even with 2,250,000 samples across 30 games, bias introduced by the embedding and sampling noise yielded unequal estimates of the win or loss probabilites of different corner squares or edge squares.

\section{Conclusions and Future Directions}

    We hope that game-playing quantum algorithms will help to test quantum computers and to make quantum computing more relatable to laypeople.  To these ends, we present an initial attempt to program a quantum computer to play a competitive, traditional game by considering all future game paths and by performing unbiased sampling of these paths in order to identify the next move that maximizes win probability. Although tic-tac-toe is a very simple game, we were pleased to find that current quantum annealers consistently defeat a random player, and force a draw most of the time against a ``perfect'' classical player when the quantum player moves first.  Win probability estimates also revealed benefits from an improved sampling strategy (inhomogeneous driving).  Extension to more complex competitive games --- nim, Connect-Four, checkers, chess, Go, \textit{etc.} --- could help to benchmark quantum hardware and quantum optimization algorithm advancements.  We also propose extending this to games of imperfect information like poker, where quantum computers may be well-suited to sampling over all future game paths given the possible distribution of current game states.  Although quantum computers theoretically offer unique potential for \textit{unbiased} sampling of the entire set of ground states (\textit{i.e.}, sampling from \textit{all} future game paths leading to a win, regardless how hard to find they may be by classical algorithms), we found that embedding issues, noise, dynamic range limits, and other practical considerations can skew sampled distributions. This makes game-playing particularly useful for testing real-world performance of quantum hardware, algorithms, and embedding strategies as they improve, to assess the extent to which less-biased quantum methods distinguish themselves from sampling path-dependent classical approaches.

\newpage

\bibliographystyle{unsrt}
\bibliography{2_QuantumTicTacToe_Bibliography}

\section*{Acknowledgements}

    \scriptsize{TZ was funded by the Natural Sciences and Engineering Research Council of Canada (NSERC) and by the Simons Foundation. VKM was funded by the Simons Foundation.  The authors thank P. Douglas Renfrew and Christoper Edelmaier for helpful conversations, and the Flatiron Institute’s Scientific Computing Core for ongoing support. The computations reported in this paper were performed using resources made available by the Flatiron Institute, or using an allocation on the D-Wave quantum annealer funded by the Simons Foundation. The Flatiron Institute is a division of the Simons Foundation.}

\section*{Conflict of Interest Statement}
    \scriptsize{VKM is a co-founder and shareholder of Menten AI, a peptide therapeutics company that uses advanced optimizers.}

% \vspace{12pt}
% \color{red}
% IEEE conference templates contain guidance text for composing and formatting conference papers. Please ensure that all template text is removed from your conference paper prior to submission to the conference. Failure to remove the template text from your paper may result in your paper not being published.

\end{document}